\documentclass[aps,prb,reprint,twocolumn]{revtex4-1}
 \usepackage{amsmath}
\usepackage{braket}
\usepackage{amssymb}
\usepackage{graphicx}
\usepackage{bm}
\usepackage{mathtools}
\usepackage{color}

\begin{document}

\title{Dimensional Crossover of Topological Edge States in Su-Schrieffer-Heeger Model}
\author{Daichi Obana,$^1$ Feng Liu,$^1$ Katsunori Wakabayashi$^{1,2}$}
\email{D.O.: 1227banao@kwansei.ac.jp,\\ F.L.: ruserzzz@gmail.com,\\ K.W.: waka@kwansei.ac.jp}
\affiliation{$^1$Department of Nanotechnology for Sustainable Energy, School of Science and Technology,
Kwansei Gakuin University, Gakuen 2-1, Sanda, Hyogo 669-1337, Japan}
\affiliation{$^2$National Institute for Materials Science (NIMS), Tsukuba, Ibaraki 305-0044, Japan}

\begin{abstract}
Su-Schrieffer-Heeger (SSH) model on two-dimensional square lattice exhibits a
 topological phase transition, which is related to the Zak phase
 determined by bulk band topology. The strong modulation of
 electron hopping causes nontrivial charge polarization even in the
 presence of inversion symmetry. The energy band structures and
 topological edge states have been calculated numerically in previous
 studies. Here, however, full energy spectrum and explicit form of wave
 functions for two-dimensional bulk and one-dimensional ribbon geometries of SSH
 model are analytically derived using wave mechanics approach. 
 Explicit analytic representations of wave functions provide the information
 of parity for each subband, localization length and critical point of topological
 phase transition in SSH ribbon.  
 It is also shown that the dimensional crossover of 
 topological transition point for SSH model from one to two-dimension.
 \end{abstract}
\maketitle

\section{Introduction}
Recent development of topological band theory in condensed matter
physics~\cite{Bansil2016,Hasan2010,Qi2011} has
established a new class of electronic materials such as 
topological insulators,~\cite{Kane2005,Bernevig2006,Fu2007,Hsieh2008,Chen2009,Chang2013,Ando2013,Sato2016}
topological crystalline
insulators,~\cite{Fu2011,Tanaka2012,Dziawa2012a,Fu2015} and topological
semimetals.~\cite{Wan2011,Burkov2011,Borisenko2014,Liang2016,Watanabe2016,kobayashi2016,Yang2017,Liu2019B}
In these topological materials, topologically protected edge states
(TESs) emerge owing to nontrivial bulk band 
topology. TESs are robust to defects and edge roughness and can be
exploited for applications to low-power consumption electronic and
spintronic devices.
One origin of TESs is nonzero Berry curvature induced by spin-orbit
couplings. Berry curvature is a geometric field strength 
in momentum space. Its integration over momentum space yields a magnetic
monopole that is characterized by the Chern number.

Even under zero Berry curvature, 
the Berry connection, a geometric vector potential whose curl yields the
Berry curvature, can also lead to 
TESs.~\cite{Liu2017} Integration of the Berry connection over momentum space (also
called the Zak phase~\cite{Zak1989}) results in an electric dipole moment that
generates robust fractional surface charges.~\cite{King1993,Resta1994,Zhou2015}
Such a dipole field related to the Zak phase is used to design 
topological materials, i.e., topological electrides~\cite{Hirayama2018,Huang2018} and A$_3$B atomic
sheet such as C$_3$N.~\cite{Liu2017B,Kameda2019} Recently, this idea is extended to an electric
quadrapole moment which induces topological corner
states.~\cite{Benalcazar2017,Benalcazar2017B,Song2017,Fukui2018,Wang2018,Ezawa2018,Liu2019,Fukui2019}
In addition, since topological design on the basis of Zak phase does not
demand the spin-orbit couplings, this approach is useful to apply to
nonelectronic systems such as topological photonic,~\cite{Liu2018,Xiao2014,Xie2018,Chen2018,Otaoptica2019} accoustic
crystals~\cite{Liuyi2017,Zhang2018,Zhang2019,Zheng2019} and topological circuit.~\cite{Liushuo2019} 

One of the most simple models to demonstrate the topological phase
transition owing to Zak phase associated with zero Berry curvature is 
Su-Schrieffer-Heeger (SSH) model\cite{Su1980,Heeger1988} on two-dimensional (2D)
square lattice.~\cite{Liu2017} 
In this model, topological phase transition occurs by tuning the ratio
between inter- and intra-cell electron hoppings.
If inter-cell hoppings 
 become larger than the
intra-cell hoppings, 
Zak phase becomes nonzero and edge states appear as a consequences of
bulk-edge correspondence. 
However, the emergence of edge states in ribbon systems has been confirmed only by numerical calculations so far.

Meanwhile, graphene is another good example that provides the finite Zak phase
accompanying TESs. Graphene has two characteristic edge structures,
i.e. zigzag and armchair. 
Zigzag graphene edges provide the robust edge localized states at the
Fermi energy,~\cite{Fujita1996,Nakada1996,Wakabayashi1999} which can be attributed to the existence
of finite Zak phase in bulk wave function of
graphene.~\cite{Delplace2011,rizzo2018,grning2018}
In actual, edge states provide the perfectly conducting
channel\cite{Wakabayashi2007,Wakabayashi2009A,Wakabayashi2009B} and
lead to very high conductivity in graphene nanoribbons.~\cite{baringhaus2014}
However, armchair graphene edges do not provide such edge states at all
owing to zero Zak phase. 
The graphene nanoribbons are particularly advantageous, since
their complete energy spectrum and wave functions can be 
analytically obtained by solving the equations of motion of
tight-binding model using wave mechanics approaches.~\cite{Wakabayashi2010,Wakabayashi.SSC.2012} 

In this paper, we analytically derive full energy spectrum and
corresponding wave functions of one-dimensional (1D) SSH ribbons using the wave mechanics approach.
From explicit form of wave functions, we obtain the information of
parity for each subbands, localization length of TES, and critical point
of topological phase transition in 1D SSH ribbons, to clarify 
crossover from 1D to 2D system.
In 2D limit, the topological phase
transition happens when the inter- and intra-cell
hoppings are equal. However, in 1D SSH ribbons, it is found that 
more stronger inter-cell hoppings are needed for topological phase
transition owing to the finite size effect. It is also found that the
critical value of transition has a power-law dependence on the ribbon
width.

The paper is organized as follows.
In Sec.~\ref{sec2}, we give a brief summary of energy spectrum and wave
functions of SSH model in 2D limit, where
the topological phase transition is related with Zak phase.
In Sec.~\ref{sec3}, we analytically derive the energy spectrum and
corresponding wave functions of SSH ribbons by using the wave mechanics approach. 
The critical ratio between intra- and inter-cell electron hoppings for
the topological phase transition is calculated by using the analytic
solutions. 
Sec.~\ref{sec4} provides the summary of the paper.

\section{2D SSH Model}\label{sec2}
\begin{figure}[ht]
  \begin{center}
   \includegraphics[width=0.45\textwidth]{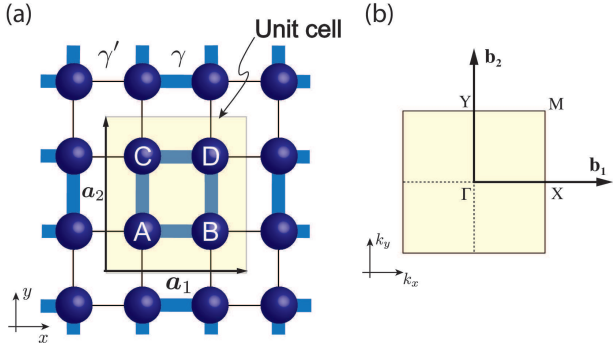}
  \end{center}
\caption{
 (a) Schematic of SSH model on square lattice. Thick and thin bonds
 represent the intra- and inter-cell electron hoppings, respectively. The primitive translation vector
 are $\bm{a_1}=(a,0)$ and $\bm{a_2}=(0,a)$. (b) Corresponding First BZ. 
The reciprocal lattice vector are $\bm{b_1}=(\frac{2\pi}{a},0)$ and $\bm{b_2}=(0,\frac{2\pi}{a})$.
}
   \label{fig:2DSSH_bulk1}
 \end{figure}

In this section, we briefly discuss the electronic states and their
topological properties of 2D SSH model. 
Figure~\ref{fig:2DSSH_bulk1}(a) shows 
schematic of 2D SSH model on square lattice. 
The yellow shaded square indicates the unit cell, in which 
there are four atomic sites labeled as $A, B, C$ and $D$.
We assume that each atomic site possesses a single electron orbital, and 
intra- and inter-cell hoppings as $-\gamma$ and
$-\gamma^\prime$, respectively. Here, $\gamma$ and $\gamma^\prime$ are
defined as positive real values.
The primitive vectors are defined as
$\bm{a_1}=(a,0)$ and $\bm{a_2}=(0,a)$, where $a$ is the lattice
constant. 
The system has $N_x$ cells along $x$-direction, and $N_y$ cells along
$y$-direction, respectively,
resulting in system size of $L_x=N_xa$ along $x$-direction, and $L_y=N_ya$ along $y$-direction.
Figure~\ref{fig:2DSSH_bulk1}(b) shows the corresponding first Brillouin zone (BZ).

The eigenvalue equation of 2D SSH model on square lattice is written as 
\begin{equation}
 \hat{H}(\bm{k})|u_j(\bm{k})\rangle = \varepsilon_j(\bm{k})|u_j(\bm{k})\rangle,
\label{eq:bulk_1}
\end{equation}
where $\bm{k}=(k_x,k_y)$ is wavenumber vector and $j(=1,2,3,4)$ is band index. 
Eigenvector is defined as 
$|u_j(\bm{k})\rangle=(\psi_{j,A}(\bm{k}), \psi_{j,B}(\bm{k}), \psi_{j,C}(\bm{k}), \psi_{j,D}(\bm{k}))^T$,
where $(\cdots)^T$ indicates the transpose of vector. 
$\psi_{j,\alpha}(\bm{k})$ ($\alpha=A,B,C,D$) is the amplitude at site $\alpha$ for $j$-th energy band at $\bm{k}$.
Hamiltonian $\hat{H}(\bm{k})$ is explicitly written as 
\begin{equation}
 \hat{H}(\bm{k})=
\left(
 \begin{array}{cccc}
  0&-\rho_x(k_x)&-\rho_y(k_y)&0 \\
  -\rho_x^*(k_x)&0&0&-\rho_y(k_y) \\
  -\rho_y^*(k_y)&0&0&-\rho_x(k_x)\\
  0&-\rho_y^*(k_y)&-\rho_x^*(k_x)&0
 \end{array} 
 \right),
\end{equation}
where $\rho_l(k_l)=\gamma+\gamma'\mathrm{e}^{ik_la} = |\rho_l(k_l)|\mathrm{e}^{i\phi_l(k_l)}$ with $l=x,y$.
Here, $\phi_l(k_l)$ is defined as the argument of $\rho_l(k_l)$ with the range of $-\pi\leq\phi_l(k_l)\leq \pi$.

By solving Eq.~(\ref{eq:bulk_1}), energy spectrum for bulk states are
obtained as
\begin{equation}
 \varepsilon_j(\bm{k})=s_1|\rho_x(k_x)|+s_2|\rho_y(k_y)|,
 \label{eq:bulk_2}
\end {equation}
where $s_1=s_2=\pm1$. The energy spectrum contains four subbands,
the relations between the band index $j$ and signs $s_1$ and $s_2$ are
summarized in Table \ref{table_1}. 
Eigenvectors for bulk states are obtained as
\begin{eqnarray}
|u_j(\bm{k})\rangle= \frac{1}{2}\left(
  \begin{array}{cccc}
  -1 \\
  s_1\mathrm{e}^{-i\phi_x(k_x)} \\
  s_2\mathrm{e}^{-i\phi_y(k_y)} \\
  -s_1 s_2\mathrm{e}^{-i\left[\phi_x(k_x)+\phi_y(k_y)\right]} \\
  \end{array}
  \right).
 \label{eq:Zak_wf}
\end{eqnarray}
 \begin{table}[t]
  \caption{Relation between band index $j$, eigenvalue $\varepsilon_j$, eigenfunction $u_j$.
  $\zeta_j$ denotes the parity of eigenvector at M, X and $\Gamma$ points for trivial (nontrivial) phase.
  }
  \label{table_1}
  \tabcolsep = 3mm
   \begin{tabular}{c|cccc||ccc}
    $j$ & $s_1$ & $s_2$ & $\varepsilon_j$ & $u_j$ & $\zeta_j$(M) & $\zeta_j$(X) & $\zeta_j$($\Gamma$) \\ \hline\hline
    $1$ &  $-$  &  $-$  & $\varepsilon_1$ & $u_1$ & $+(+)$ & $+(-)$ & $+(+)$ \\
    $2$ &  $+$  &  $-$  & $\varepsilon_2$ & $u_2$ & $-(-)$ & $-(+)$ & $-(-)$ \\
    $3$ &  $-$  &  $+$  & $\varepsilon_3$ & $u_3$ & $-(-)$ & $-(+)$ & $-(-)$ \\
    $4$ &  $+$  &  $+$  & $\varepsilon_4$ & $u_4$ & $+(+)$ & $+(-)$ & $+(+)$ 
   \end{tabular}
 \end{table}

\begin{figure}[t]
  \begin{center}
   \includegraphics[width=0.45\textwidth]{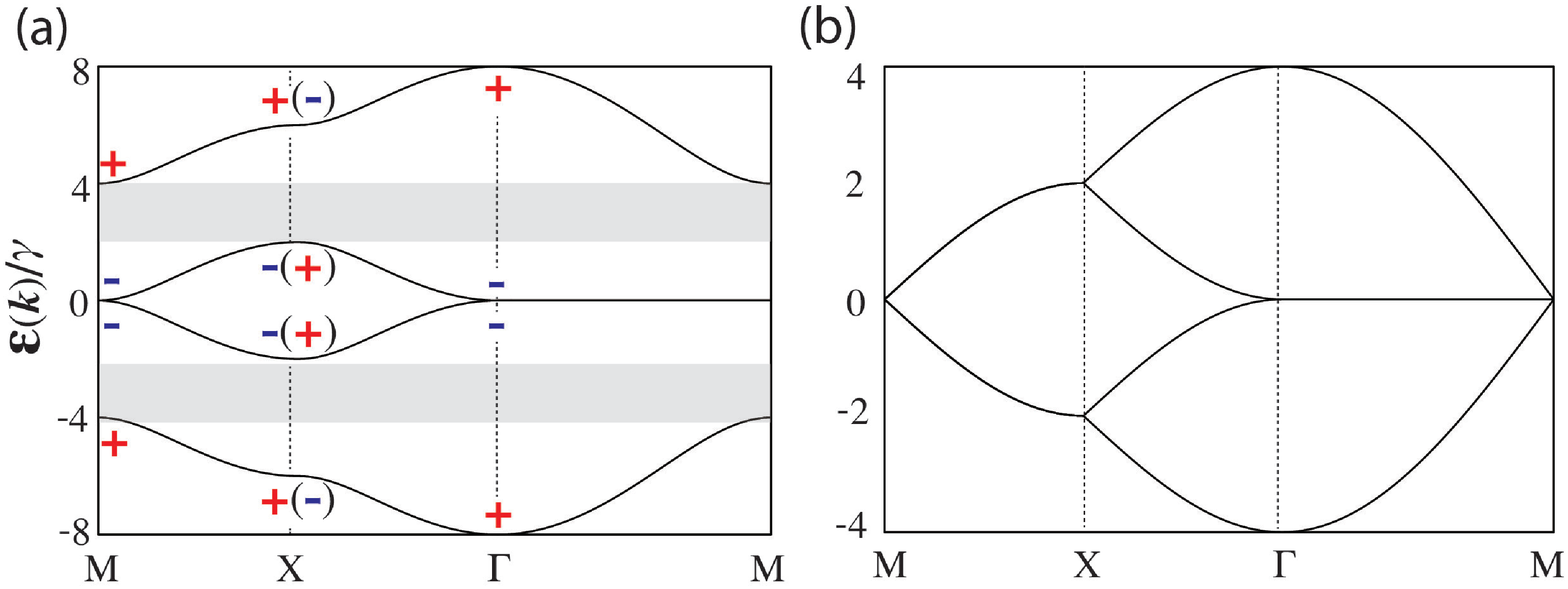}
  \end{center}
\caption{
  Energy band structure of 2D SSH model for (a) $\gamma^\prime/\gamma=1/3.0$ ($\gamma^\prime/\gamma=3.0$)
 and (b) $\gamma^\prime/\gamma=1$.
 The signs ($\pm$) in the plot of energy band structure represent the
  parity of wave function. It should be noted that band inversion occurs at X(Y) point.
}
   \label{fig:2DSSH_bulk3}  
 \end{figure}

Figures~\ref{fig:2DSSH_bulk3}(a)
and (b) show the energy band structures for $\gamma^\prime/\gamma\neq 1$
and $\gamma^\prime/\gamma= 1$, respectively.  
In general, the energy band structures of 2D SSH model become identical
between cases of the ratio $\gamma^\prime/\gamma$ and its inverse ratio $\gamma/\gamma^\prime$. 
For finite $\gamma^\prime/\gamma$, two band gaps open between 1st and 2nd subbands, and between 3rd and 4th subbands. 
The band gaps close at $\gamma^\prime/\gamma=1$. 
However, it should be noted that the parities at $\mathrm{X(Y)}$ point are inverted
between two regions of $\gamma^\prime/\gamma\le 1$ and $\gamma^\prime/\gamma> 1$.

The topological properties of 2D SSH model can be characterized in terms of
vectored Zak phase $\bm{\mathcal{Z}}=(\mathcal{Z}_x,\mathcal{Z}_y)$, which is defined as the line integration of Berry connection.
Berry connection for $j$-th energy band is defined as  
$\bm{A}_j=(a_j(k_x),a_j(k_y))$ with 
$a_j(k_l)=-i\braket{u_j(\bm{k})|\frac{\partial}{\partial k_l}|u_j(\bm{k})}$.
Since the band inversion happens in 2D SSH model at $\gamma^\prime/\gamma=1$, 
finite Zak phase appears for $|\gamma^\prime/\gamma|> 1$, 
and absent for $|\gamma^\prime/\gamma|\le 1$. 
In case of finite Zak phase, system possesses charge polarization which induces the TESs.
From now on, we call the phase with finite Zak phase {\it nontrivial} phase, otherwise {\it trivial} phase. 
The $l$-th component of vectored Zak phase in 2D SSH model can be
related with the winding phase of eigenvectors as 
\begin{eqnarray}
 \mathcal{Z}_l&=&-i
\sum_{j=1}^{occ.}
\int_0^{2\pi}\braket{u_j(\bm{k})|\frac{\partial}{\partial k_l}|u_j(\bm{k})}dk_l \nonumber \\
& = &
 N_{occ.}\frac{1}{2}\Delta\phi_l(k_l),
 \label{eq:Zak_winding}
\end{eqnarray}
where 
$\Delta\phi_l(k_l)$ is the winding phase of eigenvectors accompanied by the variation of $k_l$ from $0$ to $2\pi$
and $N_{occ.}$ is the number of occupied energy bands. 
The derivation of Eq.~(\ref{eq:Zak_winding}) is given in Appendix~\ref{apeA}. 

Figures~\ref{fig:2DSSH_bulk2} (a) and (b) show the density plot of phase $\phi_l$ in
trivial and nontrivial phases, respectively.
In the trivial case, the magnitude of phase is almost zero everywhere in momentum space. 
However, in the nontrivial case, the phases $\phi_x$ and $\phi_y$ jump by $2\pi$ along
the lines of $k_x=\pm\pi$ and $k_y=\pm\pi$, respectively. 
By looking at these figures, we can find the Zak phase as following,
\begin{eqnarray}
 \mathcal{Z}_l=\frac{1}{2}\Delta\phi_l(k_l)
                            =  \begin{cases}
                                        0   \ \ \ \ \ \ \ \ \ \ \gamma'/ \gamma \leq 1,\\
                                        \pi  \ \ \ \ \ \ \ \ \ \ \gamma'/ \gamma >1.
                                     \end{cases}
 \label{eq:Zak}
\end{eqnarray}
Thus, the nontrivial phase appears for $\gamma^\prime/\gamma>1$.
The derivation of Eq.~(\ref{eq:Zak}) is given in Appendix~\ref{apeA}. 

\begin{figure}[t]
  \centering
   \includegraphics[width=0.45\textwidth]{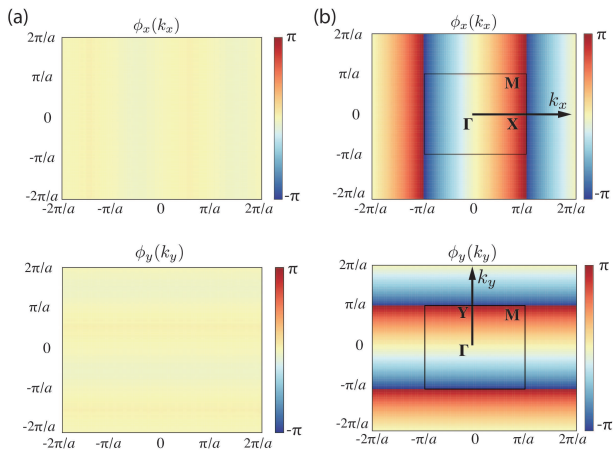}
   \caption{
 Density plot of the phase $\phi_l(k_l)$ for (a) trivial and (b) nontrivial phases.
 In trivial phase, no phase jumps occurs.
 In nontrivial phase, however, phase jumps by $2\pi$ occur 
 along the lines of $k_x=\pm\pi$ and $k_y=\pm\pi$, respectively.
   }
   \label{fig:2DSSH_bulk2}  
\end{figure}

Zak phase are related to charge polarization $\bm{P}^j=(P_x^j,P_y^j)$ as follows,~\cite{King1993,Resta1994}
\begin{equation}
P_l^j=-i\frac{1}{L_l}\sum_{j=1}^{occ.}\sum_{k_l=-\pi}^{\pi} \braket{u_j(\bm{k})|\frac{\partial}{\partial k_l}|u_j(\bm{k})}
=\frac{\mathcal{Z}_l}{2\pi}.
 \label{eq:Zak_1}
\end{equation}
Thus, if the system has the nonzero Zak phase, charge polarization
occurs, i.e. appearance of edge states.
Because of $C_4$ symmetry, 2D SSH model has ${P}_x={P}_y$ in general. 
Thus, charge polarization of 2D SSH model is $\bm{P}=(0,0)$ for trivial, and $(1/2, 1/2)$ for nontrivial phase, respectively.

 Inversion symmetry makes a strong constraint on the value of $\bm{P}$,
 which is determined gauge-independently by the parities at $\Gamma$ and X (Y) points as~\cite{Fang2012a}: 
\begin{equation}
 P_l^j=\frac{1}{2}\left(q_l^j \ \mathrm{mod} \ 2\right),
 \label{eq:Zak_2}
\end{equation}
\begin{equation}
 (-1)^{q_x^j}=\prod_{j\in occ.}\frac{\zeta_j (\mathrm{X})}{\zeta_j(\Gamma)}\ \ \ \ (-1)^{q_y^j}=\prod_{j\in occ.}\frac{\zeta_j (\mathrm{Y})}{\zeta_j(\Gamma)},
 \label{eq:Zak_3}
\end{equation}
where $P_l^j$ indicates $l$-th charge polarization, and $q_l^j$ is topological invariant which is 0 or 1.
The details of Eq.~(\ref{eq:Zak_2}) and (\ref{eq:Zak_3}) are described in Appendix~\ref{apeB}.
Thus, the value of the Zak phase determines the presence or absence of electrical polarization.

\section{Eigensystem of SSH Ribbons} \label{sec3}
In this section, we analytically derive the energy spectrum and
corresponding wave functions of 1D SSH ribbons by using wave mechanics approach under
the open boundary condition. 
This method has been successfully used to derive the energy spectrum and
wave functions for graphene nanoribbons to show the existence of edge
states.~\cite{Wakabayashi2010,Wakabayashi.SSC.2012} 
Explicit wave functions provide information of
parity for each subbands, localization length of TES.
We also inspect the topological properties of 1D SSH ribbon
and crossover from one-dimensional to 2D system.
In 2D limit, the topological phase
transition happens when the inter- and intra-cell
hoppings are equal. However, in 1D SSH ribbon, it is found that 
more stronger inter-cell hoppings are needed for topological phase
transition owing to the finite size effect. It is also found that the
critical value of transition has a power-law dependence on the ribbon
width. We also show the localization length of TES for 1D SSH ribbons
strongly depends on $\gamma^\prime/\gamma$.

\begin{figure}[ht]
  \begin{center}
   \includegraphics[width=0.45\textwidth]{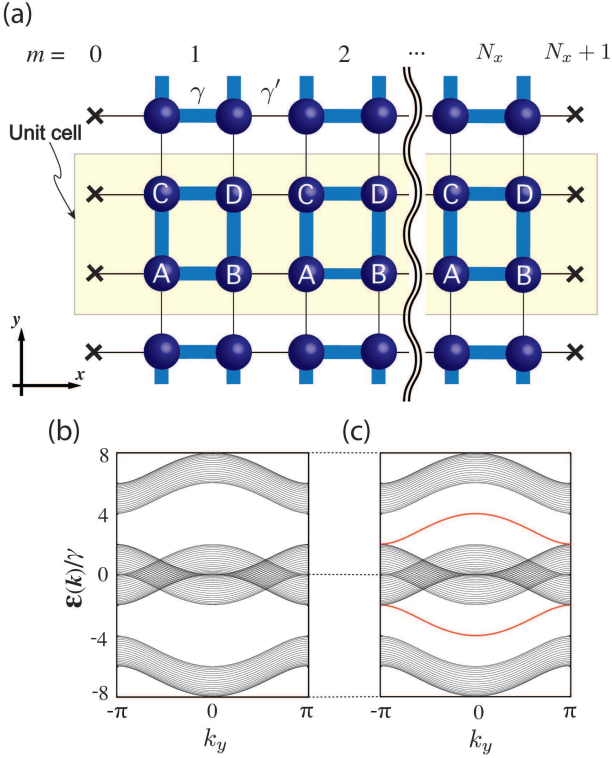}
  \end{center}
\caption{
(a) SSH ribbon model on square lattice. $N_x$ is the ribbon width. The
 $\times$ marks indicate the missing atoms for open boundary condition. 
 Energy band structure of $N_x=20$ for (b)
 $\gamma^\prime/\gamma=1/3.0$ and (c) $\gamma^\prime/\gamma=3.0$.
 In nontrivial case, edge states appear (red lines).
}
   \label{fig:2DSSH_ribbon1}  
\end{figure}

Figure~\ref{fig:2DSSH_ribbon1} shows schematic structure of SSH ribbon,
where we assume that the lattice is translational
invariant only along $y$-direction, but is finite for $x$-direction.  
From now on, we assume as $a=1$ for simplicity. 
The yellow shaded rectangle indicates unit cell of SSH ribbon,
which contains $N_x$-plaquettes, i.e. $4\times N_x$ atomic orbitals in it.  
We call four atomic sites of $m$-th plaquette as $mA$, $mB$, $mC$ and $mD$, 
and define the corresponding wave functions as $\psi_{m,A}$,
$\psi_{m,B}$, $\psi_{m,C}$ and $\psi_{m,D}$, where $m=0,1,\cdots, N_x+1$. 

The equations of motion for 1D SSH ribbon is described by 
\begin{equation}
\begin{cases}
  \varepsilon\psi_{m,A}=-\rho_y^*\psi_{m,C} -\gamma\psi_{m,B}-\gamma'\psi_{m-1,B},\\
  \varepsilon\psi_{m,B}=-\rho_y^*\psi_{m,D} -\gamma\psi_{m,A}-\gamma'\psi_{m+1,A}, \\
  \varepsilon\psi_{m,C}=-\rho_y\psi_{m,A} -\gamma\psi_{m,D}-\gamma'\psi_{m-1,D}, \\
  \varepsilon\psi_{m,D}=-\rho_y\psi_{m,B} -\gamma\psi_{m,C}-\gamma'\psi_{m+1,C}.
  \label{eq:2D_SSH_EoM}
 \end{cases} 
\end{equation}
The open boundary conditions for 1D SSH ribbon are given as
\begin{align}
 \psi_{0,B}=\psi_{0,D} = 0,
 \psi_{N_x+1,A}=\psi_{N_x+1,C} = 0.
\end{align}

Before explaining the analytic details, we shall briefly discuss the energy band structures of ribbon. 
Figures~\ref{fig:2DSSH_ribbon1} (b) and (c) show energy band structures
with $N_x=20$ in trivial phase with $\gamma^\prime/\gamma=1/3.0\le 1$,
and nontrivial phase with $\gamma^\prime/\gamma=3.0> 1$, respectively.  
These band structures have been numerically obtained by solving the equations of motion. 
In trivial phase, gapped energy band structures are obtained. However, in nontrivial phase, 
doubly degenerated TESs appear inside the gaps (indicated by magenta lines).  

Now let us derive the eigensystem of 1D SSH ribbons, 
by assuming the generic solutions of wave functions as 
\begin{equation}
  \psi_{m,\sigma}=C_{\sigma}\mathrm{e}^{i k_x m}+\tilde{C}_{\sigma}\mathrm{e}^{-i k_x m},
  \label{eq:2D_SSH_solution1}
\end{equation}
where $\sigma = A, B, C, D$. ${C}_\sigma$ and $\tilde{C}_\sigma$ are arbitrary coefficients. 
The open boundary condition leads to the following relations,
\begin{equation}
\begin{cases}
  \psi_{N_x+1,A}& = C_AZ+\tilde{C}_AZ^{-1}=0,\\
  \psi_{N_x+1,C}& = C_CZ+\tilde{C}_CZ^{-1}=0,\\
  \psi_{0,B}    & = C_B+\tilde{C}_B= 0,\\
  \psi_{0,D}    & = C_D+\tilde{C}_D= 0,
 \label{eq:2D_SSH_solution2}
\end{cases}
\end{equation}
where $Z=\mathrm{e}^{ik_x(N_x+1)}$. Thus we can rewrite the generic solution as 
\begin{equation}
 \begin{cases}
  \psi_{m,A}=C_A(\mathrm{e}^{i k_x m}-Z^2\mathrm{e}^{-i k_x m}),\\
  \psi_{m,B}=C_B(\mathrm{e}^{i k_x m}-\mathrm{e}^{-i k_x m}),\\
  \psi_{m,C}=C_C(\mathrm{e}^{i k_x m}-Z^2\mathrm{e}^{-i k_x m}),\\
  \psi_{m,D}=C_D(\mathrm{e}^{i k_x m}-\mathrm{e}^{-i k_x m}).
  \label{eq:2D_SSH_solution3}
 \end{cases} 
\end{equation}
By substituting these functions into the equations of motion, we obtain
the secular equation for 1D SSH ribbon,
\begin{equation}
 \bm{\hat{M}}\Psi = 0, \label{eq:EVprob}
\end{equation}
where $\Psi=(C_A, C_B, C_C, C_D)^T$ and $\bm{\hat{M}}$ is a $4\times 4$ matrix. The matrix elements 
$M_{i,j}$ ($i,j=1,2,3,4$) of $\bm{\hat{M}}$ are 
\begin{align}
 \begin{cases}
  M_{11}&=M_{33}=\varepsilon(\mathrm{e}^{i k_x m}-\mathrm{e}^{-i k_x m}Z^2),\\
  M_{22}&=M_{44}=\varepsilon(\mathrm{e}^{i k_x m}-\mathrm{e}^{-i k_x m}),\\
  M_{21}&=M_{43}=\rho_x\mathrm{e}^{i k_x m}-\rho_x^*\mathrm{e}^{-i k_x m}Z^2,\\
  M_{12}&=M_{34}=\rho_x^*\mathrm{e}^{i k_x m}-\rho_x\mathrm{e}^{-i k_x m},\\
  M_{13}&=\rho_y^*(\mathrm{e}^{i k_x m}-\mathrm{e}^{-i k_x m}Z^2),\\
  M_{24}&=\rho_y^*(\mathrm{e}^{i k_x m}-\mathrm{e}^{-i k_x m}),\\
  M_{31}&=\rho_y(\mathrm{e}^{i k_x m}-\mathrm{e}^{-i k_x m}Z^2),\\
  M_{42}&=\rho_y(\mathrm{e}^{i k_x m}-\mathrm{e}^{-i k_x m}),\\
  M_{14}&=M_{23}=M_{32}=M_{41}=0.
  \label{eq:ribbon_hamil_element}
 \end{cases} 
\end{align}
It should be noted that $\Psi=0$ or $k_x=0,\pm \pi$ are unphysical
solutions, because wave functions $\psi_{m,\sigma}$  
become identically zero, i.e., electrons are absent in the system. 
Thus, $\mathrm{det}\bm{\hat{M}}=0$ is demanded, and leads to the following form 
\begin{widetext}
\begin{align}
u\mathrm{e}^{4ik_xm}+v\mathrm{e}^{2ik_xm}+w+v\mathrm{e}^{-2ik_xm} Z^2+u\mathrm{e}^{-4ik_xm} Z^4=0,
\label{eq:2D_func}
\end{align}
where $u$, $v$ and $w$ are functions of $\varepsilon$, $\rho_y$,
$\gamma$, $\gamma^\prime$ and $Z$. Thus, all the coefficients of
$\mathrm{e}^{\pm i4k_x m}$, $\mathrm{e}^{\pm i2k_x m}$ terms and the
constant term should be zero to hold Eq.~(\ref{eq:EVprob}), i.e. $u=0$,
$v=0$, and $w=0$.

By using space inversion symmetry, i.e. $\psi_{N_x+1-m,A}=\pm\psi_{m,D}$, $\psi_{N_x+1-m,C}=\pm\psi_{m,B}$,
we can obtain the relations $C_D=\pm C_AZ$ and $C_B=\pm C_CZ$ from Eq.~(\ref{eq:2D_SSH_solution2}). 
In addition, the mirror symmetry leads to $\psi_{m,A}=\pm\psi_{m,C}$, i.e., $C_C=\pm C_A$.
Thus, the general form of the wave function can be written as 
\begin{equation}
 \left(
  \begin{array}{cccc}
  \psi_{m,A} \\
  \psi_{m,B} \\
  \psi_{m,C} \\
  \psi_{m,D} \\
  \end{array}
  \right)=N_c
   \left(
  \begin{array}{cccc}
  \mathrm{sin}[k_x(N_x+1-m)] \\
  (-1)^{r} s_1\ \mathrm{sin}[k_x m] \\
  s_2\ \mathrm{sin}[k_x(N_x+1-m)] \\
  (-1)^{r} s_1 s_2\ \mathrm{sin}[k_x m] \\
  \end{array}
  \right),
   \label{eq:ribbon_wf}
\end{equation}
where $r=1,2,3,\cdots, N_x$ indicates the band index of 1D SSH
 ribbon. However, as it can be clarified later, the index will be
 $r=1,2,3,\cdots, N_x-1$ in nontrivial phase, because the one missing
 mode will form the mode of TESs.
It should be noted that the parity of wave function clearly depends on
 the band index $r$. 
$N_c$ is normalization constant. 
Owing to the translational invariance along $y$-direction, 
the wave function for whole ribbon
 system can be obtained by multiplying the Bloch phase 
${\rm e}^{ik_y y}$ and Eq.~(\ref{eq:ribbon_wf}).

The coefficients of $\mathrm{e}^{\pm i4k_xm}$ are shown to be identically
 zero by using the bulk energy spectrum Eq.~(\ref{eq:bulk_2}), i.e.
\begin{align}
u(k_x, k_y,N_x)
& =\varepsilon^4-2\varepsilon^2(|\rho_x|^2+|\rho_y|^2)^2+(|\rho_x|^2-|\rho_y|^2)^2\nonumber \\
&  =\{ \varepsilon^2-(|\rho_x|\pm|\rho_y|)^2 \}^2
  =0. \label{eq:u_func} 
\end{align}
Thus, $u(k_x, k_y, N_x)$ is irrelevant for later discussion.

The solutions of $v=w=0$ determine the transverse wave number $k_x$ for
a given $\gamma^\prime/\gamma$. Thus, we obtain 
\begin{align}
v(k_x,k_y,N_x)&
=\varepsilon^4[Z+Z^{-1}] \nonumber \\
&-\varepsilon^2[(\rho_x^{*2}+|\rho_x|^2+2|\rho_y|^2)Z+(\rho_x^{2}+|\rho_x|^2+2|\rho_y|^2)Z^{-1}]\nonumber \\
&+(|\rho_x|^2-|\rho_y|^2)[(\rho_x^{*2}-|\rho_y|^2)Z+(\rho_x^{2}-|\rho_y|^2)Z^{-1}]=0,\label{eq:v_func} \\
w(k_x,k_y,N_x)&=\varepsilon^4[Z^2+Z^{-2}+4] \nonumber \\
&-2\varepsilon^2[(\rho_x^{*2}+|\rho_y|^2)Z^2+(\rho_x^{2}+|\rho_y|^2)Z^{-2}
                                                         +(\rho_x+\rho_x^*)^2+4|\rho_y|^2]\nonumber \\
&+(\rho_x^{*2}-|\rho_y|^2)^2Z^2+(\rho_x^{2}-|\rho_y|^2)^2Z^{-2}
                                   +4\rho_x^2\rho_x^{*2}-2|\rho_y|^2(\rho_x+\rho_x^*)^2+4|\rho_y|^4=0.
\label{eq:w_func}
\end{align}
\end{widetext}
It should be noted that both $v(k_x,k_y,N_x)$ and $w(k_x,k_y,N_x)$ are periodic
and even functions of $k_x$ with a period of $2\pi$, i.e.
$v(k_x,k_y,N_x)=v(-k_x,k_y,N_x)$ and $v(k_x+2n\pi,k_y,N_x)=v(k_x,k_y,N_x)$ ($n$ is
arbitrary integer). Similarly, $w(k_x,k_y,N_x)$ has the properties of 
$w(k_x,k_y,N_x)=w(-k_x,k_y,N_x)$ and $w(k_x+2n\pi,k_y,N_x)=w(k_x,k_y,N_x)$. 
Thus, it is sufficient to find the solutions within the range $0<k_x<\pi$.
\begin{figure*}[ht]
  \centering
   \includegraphics[width=0.95\textwidth]{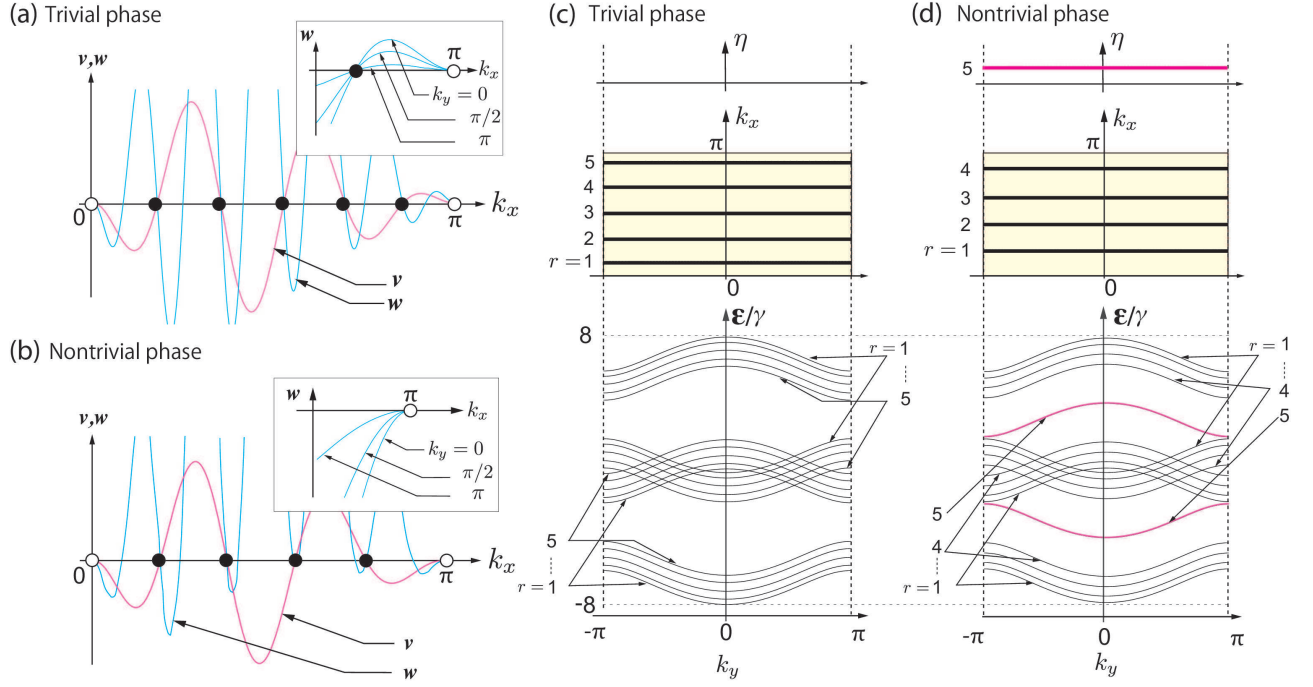}
   \caption{
 Schematic chart of function, $v$ (magenta line) and $w$ (cyan line) 
for $N_x=5$ in (a) trivial [$|\gamma^\prime/\gamma| \leq (\gamma^\prime/ \gamma)_c$] and
 (b) nontrivial [$|\gamma^\prime/\gamma| > (\gamma^\prime/
 \gamma)_c$] phases, respectively.  
White circles at $k_x=0$ and $\pi$ are
 unphysical solutions. 
Black circle in the region of $0<k_x<\pi$ means real value solution of
 transverse wavenumber $k_x$. 
It should be noted that there are $N_x$ solutions 
 for trivial phase, but only $N_x-1$ solutions for
 nontrivial phase. The insets show dependence of $w$ on 
 longitudinal wave number $k_y$ near $k_x=\pi$. 
 The relation between obtained complex transverse wavenumber $k_x+i\eta$
 and corresponding energy band structure for 1D SSH model with $N_x=5$
 for (c) trivial and (d) nontrivial phases. 
 In the upper panels of (c) and (d),
 solutions of complex transverse
 wavenumber $k_x+i\eta$ in BZ are plotted. Both of $k_x$ and $\eta$ do
 not have $k_y$ dependence, and no imaginary part exists in trivial phase.
 The lower panels of (c) and (d), the corresponding energy band
 structures.
 Real value solutions ($k_x$) lead to energy dispersion of bulk states
 denoted by thin black lines, however, imaginary solution ($\eta$) leads
 to energy dispersion of TESs denoted by thin magenta lines.
   }
 \label{fig:2DSSH_ribbon3}
\end{figure*}

At $\gamma^\prime/\gamma=1$, 
Eqs.~(\ref{eq:v_func}) and (\ref{eq:w_func}) safely reproduce 
the energy spectrum of simple 2D square lattice, i.e. 
$\varepsilon(\bm{k}) = \pm 2\gamma\left[\cos\left(\frac{k_x}{2}\right)\pm\cos\left(\frac{k_y}{2}\right)\right]$.
In addition, 
the open boundary condition for ribbon structure leads to the transverse wave numbers $k_x$ as
\begin{eqnarray}
k_x=\frac{m}{N_x+1}\pi,\ \ \ \ \ m=1,2,3,\cdots,N_x. 
\end{eqnarray}

Figures~\ref{fig:2DSSH_ribbon3}(a) and (b) show the $k_x$ dependence of 
$u(k_x,k_y,N_x)$, $v(k_x,k_y,N_x)$ and $w(k_x,k_y,N_x)$ for trivial and nontrivial
phases, respectively. Here, we have fixed the ribbon width as $N_x=5$. 
The functions $u(k_x,k_y,N_x)$, $v(k_x,k_y,N_x)$ and $w(k_x,k_y,N_x)$ are calculated
under the condition of $\varepsilon^2=(|\rho_x|+|\rho_y|)^2$.
The case of $\varepsilon^2=(|\rho_x|-|\rho_y|)^2$ is not shown, 
because the results do not change. The solutions of 
$u(k_x,k_y,N_x)$, $v(k_x,k_y,N_x)$ and $w(k_x,k_y,N_x)$,
which are denoted by black circles in Figs.~\ref{fig:2DSSH_ribbon3} (a)
and (b), give the transverse wavenumbers $k_x$ which determine the
eigenstates of 1D SSH ribbon. 
The variation of longitudinal wavenumber $k_y$ only alters the amplitude
of $u(k_x,k_y,N_x)$, $v(k_x,k_y,N_x)$ and $w(k_x,k_y,N_x)$, and does not change the
positions of zero points. 
Thus, the solutions of $\mathrm{det}\bm{M}=0$ do not depend on $k_y$.

For a fixed $N_x$, the number of real solutions $N$ for
$\mathrm{det}\bm{\hat{M}}=0$ becomes different depending on the value
of $\gamma^\prime/\gamma$, i.e.
\begin{eqnarray}
N=
 \begin{cases}
  N_x,   \ \ \ \ \ \ \ \ \ \ |\gamma^\prime/ \gamma|\leq(\gamma^\prime/ \gamma)_c\\
  N_x-1,\ \ \ \ \ |\gamma^\prime/ \gamma|>(\gamma^\prime/ \gamma)_c.
 \end{cases}
\label{eq:w_func3}
\end{eqnarray}
Here, $(\gamma'/ \gamma)_c$ is the critical value at which topological phase transition occurs.
$(\gamma'/ \gamma)_c$ can be derived by
\begin{eqnarray}
 \left. \frac{\partial}{\partial k_x}v(k_x,k_y,N_x)\right|_{k_x=\pi}=0.
\label{eq:critical}
\end{eqnarray}

In further, as indicated by Eq.~(\ref{eq:w_func3}), one real solution is
missing for $|\gamma^\prime/ \gamma|>(\gamma^\prime/ \gamma)_c$.
This disappearance of real solution happens near $k_x=\pi$ by looking at
$w(k_x,k_y,N_x)$ under the evolution of $\gamma^\prime/\gamma$, as shown
in insets of Figs.~\ref{fig:2DSSH_ribbon3} (a) and (b).
Thus, the remaining missing solution can be obtained by analytical
continuation of $k_x \rightarrow \pi + i\eta$. 
This imaginary solution $\eta$ depends on $\gamma^\prime/\gamma$.
If $\gamma^\prime/\gamma$ is larger, $\eta$ also increases.

Figures~\ref{fig:2DSSH_ribbon3} (c) and (d) show
the relation between the obtained transverse complex wavenumber
($k_x+i\eta$) and energy band structures of 1D SSH ribbons with $N_x=5$
for trivial and nontrivial phases, respectively. 
In trivial phase as shown in Fig.~\ref{fig:2DSSH_ribbon3} (c), no
imaginary solutions appears. The real transverse wavenumbers $k_x$
(denoted by thick black lines) lead to the energy subbands of extended bulk states.
However, in trivial phase as shown in Fig.~\ref{fig:2DSSH_ribbon3} (d), 
an imaginary solution ($\eta$) appears, which gives the subband of TESs (thin magenta lines).
Note that edge states are doubly degenerate owing to inversion symmetry.

\begin{figure}[ht]
  \centering
   \includegraphics[width=0.45\textwidth]{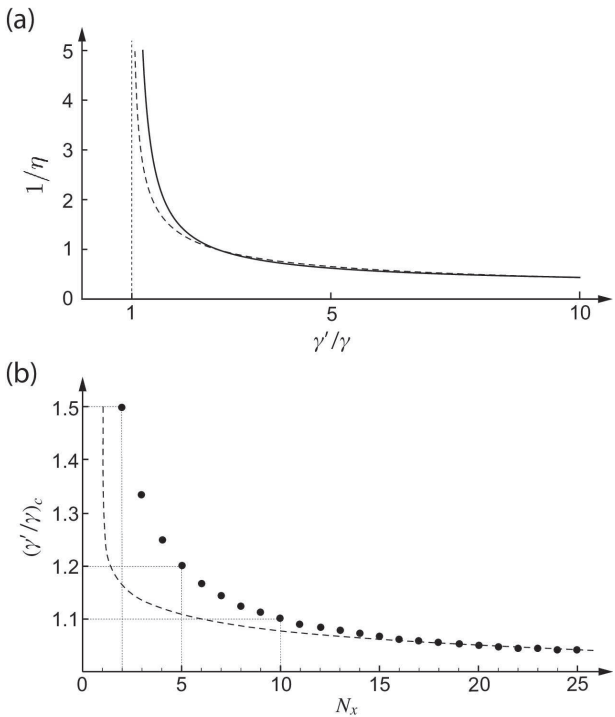}
 \caption{
(a) Dependence of localization length $1/\eta$ of TES on
 $\gamma^\prime/\gamma$.
Solid line is calculated from analytic wave functions of
 Eq.~(\ref{eq:ribbon_wf_TES}).
Dashed line is the fitting for the large coupling limit. 
The fitting curve is $1/\eta=1.327(\gamma^\prime/\gamma-1)^{-0.481}$. 
(b) Ribbon width ($N_x$) dependence of critical value
 $(\gamma^\prime/\gamma)_c$, where topological phase transition occurs. 
In 2D limit of $N_x\rightarrow\infty$, $(\gamma^\prime/ \gamma)_c$ converges to $1$.
Black circles are $(\gamma^\prime/\gamma)_c$ obtained from
 Eq.~(\ref{eq:critical}).
Dashed line is the fitting for the limit of large $N_x$.
The fitting curve is $(\gamma^\prime/\gamma)_c=1.165(N_x-1)^{-0.034}$.  
}
   \label{fig_6}  
\end{figure}

The wave function of TESs can be written as 
\begin{equation}
 \left(
  \begin{array}{cccc}
  \psi_{m,A} \\
  \psi_{m,B} \\
  \psi_{m,C} \\
  \psi_{m,D} \\
  \end{array}
  \right)=N_c
   \left(
  \begin{array}{cccc}
  \mathrm{sinh}[\eta (N_x+1-m)] \\
  (-1)^{N_x} s_1\ \mathrm{sinh}[\eta m] \\
  s_2\ \mathrm{sinh}[\eta(N_x+1-m)] \\
  (-1)^{N_x} s_1 s_2\ \mathrm{sinh}[\eta m] \\
  \end{array}
  \right),
   \label{eq:ribbon_wf_TES}
\end{equation}
where $N_c$ is normalization constant. 
Similar to the extended states, the wave function of TESs for whole ribbon
 system can be obtained by multiplying the Bloch phase 
${\rm e}^{ik_y y}$ and Eq.~(\ref{eq:ribbon_wf_TES}). 
It should be noted that the parity of TESs depends on $N_x$, resulting in
the even-odd effect of $N_x$ on parity. This property can be used for
band-selective filter.~\cite{Nakabayashi2009}  
We should note that the wave
function for TESs contains hyperbolic sine functions which are
characterized by the imaginary transverse wavenumber $\eta$. Thus, TESs 
are localized states with the characteristic localization length of $1/\eta$. 

Figure~\ref{fig_6} (a) shows the electron hopping
($\gamma^\prime/\gamma$) dependence of the localization length
($1/\eta$).
$1/\eta$ decays with the power-law
on increase of $\gamma^\prime/\gamma$. 
With increase of $\gamma^\prime/\gamma$, the localization
length of TESs becomes shorter, i.e. electrons are more strongly
localized near the edges of ribbon. 
Thus, if the localization length is larger than ribbon width, the
destructive interference between TESs from both edges occurs. 
In actual, this property of TESs demands that more stronger 
$\gamma^\prime/\gamma$ is necessary to induce the topological phase
transition if the ribbon width gets narrower. 
Figure~\ref{fig_6} (b) shows the ribbon width ($N_x$) dependence of 
$(\gamma^\prime/\gamma)_c$, which is the critical value for topological
phase transition. 
Though $(\gamma^\prime/\gamma)_c=1$ for 2D SSH model,
Figure~\ref{fig_6} (b) shows that narrower SSH ribbons have larger 
$(\gamma^\prime/\gamma)_c$ than 1. Since 
$(\gamma^\prime/\gamma)_c$ slowly decays with increase of $N_x$ owing to
its power-law behavior, the finite size effect persists even in the very
large $N_x$. Thus, finite size effect for topological phase transition in 1D SSH 
model becomes crucial especially for narrower ribbons. 

\section{summary}\label{sec4}
In summary, we have analytically derived eigensystems of 2D SSH and 1D SSH ribbon
models on square lattice by using wave mechanics approach.
In these models, the modulation of electron hopping causes
nontrivial charge polarization even in the presence of inversion symmetry. 
As for 2D SSH model, it has been known that the topological phase transition
occurs, accompanied by nontrivial charge polarization 
if inter-cell hopping $\gamma^\prime$ are larger than intra-cell
hopping $\gamma$. Owing to bulk-edge correspondence, TES are expected to
appear in the nontrivial phase. However, it has been confirmed only by
numerical calculations so far. 
In this paper, we have successfully derived full energy spectrum and
corresponding wave functions for 1D SSH ribbons by using the wave
mechanics approach. Mathematical relation between complex transverse
wavenumber ($k_x+i\eta$) and energy band structures of 1D SSH ribbon are
clarified. Also, the parity of wave functions and localization length of TESs
are analytically identified. 

It is also found that the critical value of
topological phase transition $(\gamma^\prime/\gamma)_c$ strongly depends
on the ribbon width and deviates from 1 which is the limit of 2D SSH
model. This information will be necessary when we try to fabricate the
topological electronic, photonic devices in quasi-1D geometry on the
basis of SSH model. 

By using the wave function used in this paper, we can discuss further
electronic states of 2D SSH model subjected to other perturbations.
Also, our approach is useful for constructing of Green's functions by 
the decomposition of propagating and evanescent modes, which is need for
atomistic calculations of electronic transport
properties.~\cite{Deng2014}  
Our results will serve to design new 2D materials which possess non-zero
Zak phase and edge states which are necessary for robust electronic
transport. Furthermore, it can be applied to theory of topological
photonic crystals.

F.L. is an overseas researcher under the Postdoctoral Fellowship of the
Japan Society for the Promotion of Science (JSPS). K.W. acknowledges
the financial support from Masuya Memorial Research Foundation of
Fundamental Research. This work was supported by JSPS KAKENHI Grants
No. JP17F17326, and No. JP18H01154.

\appendix
\section{Derivation of Eqs. (\ref{eq:Zak_winding}) and (\ref{eq:Zak})}\label{apeA}
Here we derive the relation between Zak phase and winding phase of eigenvectors given in 
Eqs.~(\ref{eq:Zak_winding}) and (\ref{eq:Zak}).
From eigenvector of Eq.~(\ref{eq:Zak_wf}), we obtain
\begin{align}
\braket{u_j(\bm{k})|\frac{\partial}{\partial k_l}|u_j(\bm{k})} = \frac{i}{2}\frac{d\phi_l(k_l)}{dk_l}.
\end{align}
Thus,
Eq.~(\ref{eq:Zak_winding}) can be derived as
\begin{eqnarray}
 \mathcal{Z}_l&=& - i \sum_{j=1}^{occ.}\int_{0}^{2\pi}
\braket{u_j(\bm{k})|\frac{\partial}{\partial k_l}|u_j(\bm{k})}
dk_l \nonumber\\
&=&N_{occ.}\frac{1}{2}\int_0^{2\pi} dk_l\frac{d\phi_l(k_l)}{dk_l}\nonumber\\ 
&=&N_{occ.}\frac{1}{2}\oint d\phi_l(k_l)\nonumber\\ 
&\eqqcolon&N_{occ.}\frac{1}{2}\Delta\phi_l(k_l),
 \label{eq:ape1}
\end{eqnarray}
where $\Delta\phi_l(k_l)$ is the winding phase of eigenvectors accompanied by the variation of $k_l$ from $0$ to $2\pi$ 
and $N_{occ.}$ is the number of occupied energy bands. 

From $\rho_l(k_l)=|\rho_l(k_l)|\mathrm{e}^{i\phi_l(k_l)}$,
total derivative of $\rho_l(k_l)$ is obtained as
\begin{align}
 d\rho_l(k_l)=\mathrm{e}^{i\phi_l(k_l)}d|\rho_l(k_l)|+i|\rho_l(k_l)|\mathrm{e}^{i\phi_l(k_l)}d\phi_l(k_l).
\end{align}
Identically, we have 
\begin{align}
 \frac{1}{\rho_l(k_l)}d\rho_l(k_l)=\frac{1}{|\rho_l(k_l)|}d|\rho_l(k_l)|+id\phi_l(k_l).
\end{align}
By taking the contour integration for both sides, we obtain the winding phase of eigenvectors as 
\begin{align}
 \frac{1}{i}\oint\frac{1}{\rho_l(k_l)}d\rho_l(k_l)=\Delta\phi_l(k_l),
\end{align}
where $\rho_l(k_l)=\gamma^\prime/\gamma+\mathrm{e}^{ik_l}$. 
We should note that the integrand of the left hand side has a singular point at the origin. 
The path of the contour integration is a unit circle at the center $\gamma^\prime/\gamma$ in the complex plane, which is parameterized by $k_l$.
Thus, we can evaluate the value of $\Delta \phi_l$ using residue theorem.

In trivial phase, i.e. $\gamma^\prime/\gamma\leq 1$, $\Delta \phi_l=0$,
because the contour does not enclose the origin. In nontrivial phase,
i.e. $\gamma^\prime/\gamma>1$, however, $\Delta \phi_l=2\pi$,  
because contour encloses the origin. 
Therefore, Zak phase in 2D SSH model is given as Eq.~(\ref{eq:Zak}).

\section{Derivation of Eqs. (\ref{eq:Zak_2}) and (\ref{eq:Zak_3})}\label{apeB}
The charge polarization can be related with the parity of wave functions. 
In order to show the relation, let us define the sewing matrix $\omega$, which is in form of the space
inversion operator $\Pi$ for formulation of Zak phase for multi-band system.
Here, $\Pi$ an operator that inverts position as $\Pi u(\bm{k})=u(\bm{-k})$. 
The sewing matrix $\omega$ is define as
\begin{eqnarray}
\omega_{ij}(\bm{k})\coloneqq \braket{u_i(-\bm{k})|\Pi|u_j(\bm{k})},
 \label{eq:apeC_1}
\end{eqnarray}
which $i,j$ are band index. In addition, we define 
Berry connection matrix as
$\bm{a}_{ij}(\bm{k})=-i\braket{u_i(\bm{k})|\nabla_{\bm{k}}|u_j(\bm{k})}$. 
By using this matrix, Eq.~(\ref{eq:Zak_1}) can be rewritten as 
\begin{eqnarray}
 P_l&=&\sum_{j=1}^{occ.}\int_{-\pi}^{\pi} \frac{dk_l}{2\pi}a_{jj}(k_l) \nonumber\\
  &=&\frac{1}{2\pi}\int_{-\pi}^{\pi}dk_l\ \mathrm{Tr}[\bm{a}_{ij}(k_l)].
 \label{eq:apeC_2}
\end{eqnarray}

In the presence of space inversion symmetry,
the integrand of Eq.~(\ref{eq:apeC_2}) can be rewritten as
\begin{eqnarray*}
\mathrm{Tr}[\bm{a}_{ij}(\bm{k})]=\mathrm{Tr}[\bm{a}_{ij}(-\bm{k})]+i\mathrm{Tr}[\omega_{ij}^\dagger(\bm{k})\nabla_{\bm{k}}\omega_{ij}(\bm{k})].
\end{eqnarray*}
Therefore, charge polarization is expressed as
\begin{eqnarray*}
 P_l&=&\frac{1}{2\pi}\int_{-\pi}^{\pi}dk_l \ \left\{ \mathrm{Tr}[\bm{a}(-\bm{k})]+i\mathrm{Tr}[\omega^{\dag}(\bm{k})\frac{\partial}{\partial k_l}\omega(\bm{k})]\right\}\\
 &=&-P_l+q_l^j.
\end{eqnarray*}
Here, $q_l^j$ is right hand of integral element. 
Using the unitarity of $\omega(\bm{k})$, the integrand of $q_x^j$ can be
rewritten as 
$\mathrm{Tr}[\omega^\dag(\bm{k})\nabla_k
\omega(\bm{k})]=\nabla_k\mathrm{log}(\mathrm{det}[\omega(\bm{k})])$.
Thus, we obtain
\begin{eqnarray}
 q_x^j&=&-\frac{i}{2\pi}\int_{-\pi}^{\pi}dk_l\frac{d\mathrm{ln}(\mathrm{det}[\omega(k_x,0)])}{dk_l} \nonumber\\
 &=&-\frac{i}{\pi}\mathrm{ln}\left\{ \frac{\mathrm{det}[\omega(\pi,0)]}{\mathrm{det}[\omega(0,0)]} \right\}.
 \label{eq:apeC_q1}
\end{eqnarray}
Since eigenvalue of space inversion operator $\Pi$ is $\pm1$,
$\mathrm{det}[\omega(\bm{k})]$ is given as 
\begin{eqnarray}
 \mathrm{det}[\omega(\bm{k})]=\prod_{j\in occ.}\zeta_j(\bm{k}),
 \label{eq:apeC_q2}
\end{eqnarray}
where $\zeta_j(\bm{k})$ are the  parity of wave function in $j$-th energy
band at $\bm{k}$. Substituting Eq.~(\ref{eq:apeC_q2}) into Eq.~(\ref{eq:apeC_q1})
leads to 
\begin{align}
 \mathrm{ln}(\mathrm{e}^{i\pi q_x^j})&=
\mathrm{ln}
\left\{ \prod_{j\in occ.}\frac{\zeta_j(\pi,0)}{\zeta_j(0,0)}
\right\},
\nonumber \\
 (-1)^{q_x^j}&= \prod_{j\in occ.}\frac{\zeta_j(\mathrm{X})}{\zeta_j(\Gamma)}.
\label{eq:apeC_q3}
\end{align}
Similarly, $y$ component is also obtained by replacing $\mathrm{X}$ with $\mathrm{Y}$. 
Thus $q_l^j$ is topological invariant, which gives either 0 or 1.

\bibliographystyle{apsrev4-1} 
\bibliography{reference}

\end{document}